\documentclass[12pt]{article}
\usepackage{graphics,epsfig}

\begin{document}

\title{Global Warming: the Sacrificial  Temptation}

\author{Serge Galam\thanks{E-mail: serge.galam@polytechnique.edu}\\
Centre de Recherche en \'Epist\'emologie Appliqu\'ee (CREA), \\\
\'Ecole Polytechnique and CNRS, \\
1 Rue Descartes, 75005 Paris, France }

\date{August 30, 2007}

\maketitle

\begin{abstract}
The claimed unanimity of the scientific  community about the human culpability for global warming is questioned. Up today there exists no scientific proof of human culpability. It is not the number of authors of a paper, which validates its scientific content. The use of probability to assert the degree of certainty with respect the global warming problem is shown to be misleading. The debate about global warming has taken on emotional tones driven by passion and irrationality while it should be a scientific debate. The degree of hostility used to mull any dissonance voice demonstrates that the current debate has acquired a quasi-religious nature. Scientists are behaving as priests in their will ``to save the planet". We are facing a dangerous social phenomenon, which must be addressed from the social point of view. The current unanimity of citizens, scientists, journalists, intellectuals and politicians is intrinsically worrying. The calls to sacrifice our way of life to calm down the upset nature is an emotional ancestral reminiscence of archaic fears, which should be analyzed as such. 
\end{abstract}

\section{The current situation: man is guilty}

Global warming has become a world issue and everyone everywhere is dealing with it, either directly, with solid devastations and human lost, or by strong disturbances in the normal local weather conditions. The media are making their daily headlines with the latest catastrophes, and politicians are addressing the issue in their talks. But yet not much has been done. It is becoming a world political shake. Pressure is growing from the public opinion and merely from non-governmental organizations to take drastic measures against the well-identified cause of the global warming. 

The scientific community is extremely active on the issue by setting detailed scenarios on the dramatic consequences of the current trend and urge governments to act immediately. The Intergovernmental Panel on Climate Change (IPCC) is monitoring a world activity with thousands of climatologists involved. They are talking with a unique and single voice about the scientific diagnostic. During their last meeting in Paris in February 2007 they concluded unanimously that it is the increased quantity of carbon dioxide in the atmosphere, which produces the global warming, and they designate man as the cause of it.

Human greed, by its exponential appetite of natural resources, is destroying the planet in pure wastes. At present rate of carbon dioxide production, global warming will lead to a total catastrophe.  Artists are getting involved in this survival cause and Al Gore is leading a new crusade to save the planet. Huge free concerts are taking places worldwide and demonstrations are organized locally. 

While American President G. W Bush has been reticent to adopt counter measures, European leaders are taking the lead to the carbon dioxide reduction, in particular the German Chancellor A. Merkel and the new French President N. Sarkozy. European Union has decided on unilateral twenty percent cut of carbon dioxide production by 2020. Many industries are taking this coming restriction very seriously and are trying to adapt without too much damage to their development. The fact that countries like China and India are not yet engaged on that reduction line, on the contrary, they are increasing their carbon dioxide production enormously, does not worry the western hard liners of immediate cuts.

To sum up the current situation, a world danger has been clearly identified by scientists, its cause determined precisely and the solution is clearly set. Man is guilty and must pay the price. Globalization and the consumerist way of life of modern society must be sacrificed on the autel of man redemption to save our beloved martyr planet. It is in the name of Science, it is in the name of human humility. We are presenting an up-side view with an emphasis on the social mechanisms, which drive the current hysteria [1, 2, 3, 4]. It is not an exhaustive review of the problem but the presentation of a specific focus [5].

\section{Something sounds wrong in this so clear picture}

The current unanimity of citizens, scientists, journalists, intellectuals and politicians is intrinsically worrying in particular since it is underlined by the generalized fuzzy feeling that man went to far in intervening in all natural processes at its single immediate profit. The clear identification of a unique culprit, man abuses, is too much reminiscent of past ancestral reactions to collective fears provoked by natural devastations. The designation of the US as the bad guy is also problematic within the current world political frame of anti Bush attitude. It appears that several different concerns all coalesce on the designation of a kind of world devil responsible of all the evils.

More problematic is the insistence of presenting the unanimity of the scientific world behind the identification of the programmed world catastrophe driven by man production of carbon dioxide. Claiming that thousands of scientists have voted like a single man to assert the certainty, estimated at $90\%$ that humans are responsible of all the observed climatologic deregulations hide something. In particular when the sole grounds of these affirmations are essentially based on models simulated with computers. The major real fact being the observed correlation between on the one hand the increase of carbon dioxide in the atmosphere, and on the other hand the measured of some increase in global temperatures. But an observed correlation does not mean a single directed relation of a cause to an effect.

\section{Remembering the past collective fears of human societies}

In the past of human history, the identification of a single responsible of all the difficulties and hardships of a society has always produced huge human destructions. Nothing good has never ever emerged from such unanimity of all parts of a society, quite the contrary. When the collective fears are driven against a designated culprit and monitored by a big political design for a better society, the result has been always fascism and desolation.

The collective human society is exposing itself to great risks in this voluntary and headlong rush to "salvation", with the real possibility of destroying itself even before the global mean temperature has had time to rise significantly.  

But why to provoke an anxiety about the proposed measures against global warming by creating such a dissonance about the apparent reassurance of official concern about a possible impending catastrophe?

\section{When a scientific matter becomes emotional}

The unanimity exhibited everywhere is indeed obtained by the exclusion of any person who dares to cast a doubt about the man guilt truth. Verbal and written aggressions as well funds cut are immediately applied against those few sceptics. In particular the much-trumpeted unanimity of the climatologist community has been obtained by the expedient of excluding those sceptical colleagues.

The debate about global warming has taken on such emotional tones that what was originally a scientific debate has now become transformed into a phase typified by passion and irrationality. The degree of hostility used to mull any dissonance voice demonstrates that the current debate has acquired a quasi-religious nature and thus has become extremely dangerous. 

To give an illustration about what should be a scientific debate, imagine that a scientist was to question the reality of gravity. The community of scientists as a whole would be indifferent to such an opinion ad just ignore it. And its scientific colleagues would at the very most express some feeling of compassion toward someone who had evidently lost his head.  But with certainty this scientist would not become the victim of any kind of aggression. 

The violence actually exercised against scientists sceptical about the cause of global warming is an additional signÑwere any indeed neededÑthat the "official" thesis of human guilt has an extremely shaky foundation. I myself went through virulent attacks from many different sources after I published a paper in the daily French newspaper Le Monde, in which I stated there exists scientific unanimity about human responsibility in global warming but no proved scientific certainty.

\section{The ancestral temptation of sacrifice}

The consensual solution embodied by this assertion of human guilt is very reassuring against the archaic fears against human vulnerability to natural elements. First it identifies the cause of the threat without uncertainty. Second it offers the clear solution to resolve the problem and suppress the threat. Third it requires sacrificing the current standard of life, which for many, is synonymous of exaggerations and abuses.

Moreover it subscribes to our historical records. Throughout history, it is found that our ancestors while facing unchained natural elements had the tendency to persuade themselves that they were the cause of it. Always they associated big and small natural catastrophes to God anger against mankind sins. God was upset and exerted the deserved punishment through the violence of nature. And for many millennia, human beings believed that they could stop this violence by a redemption marked by animal and human sacrifices. 

Fortunately, the growth of scientific understanding has taught us that there was no foundation to this custom. And yet all of a sudden, against all expectations this ancient and archaic system of beliefs is resurgent at once with fresh vitality. The incredible paradox is that scientists in the name of science monitor it.

And just as in ancient times, the new prophets are announcing the end of the world. Again as formerly, it is our greedy and profligate ways of life that are responsible for this imminent end. And again the prophets are demanding sacrifices in order to pacify nature.  Fortunately this time, they are not demanding that we sacrifice our lives, but instead that we sacrifice our way of life, including technological progress and scientific research.

\section{There exists at present no scientific certainty about human guilt}

To embody the various aspects of the global warming debate it is essential to come back to the supposed certainty of the scientific proof stating man is guilty. All media and journals assert the scientific proof by quoting especially the 2007 UNESCO February meeting of the GICC hold in Paris where 2500 scientists voted in favour of the human guilt.

Here stands a major confusion between what is a political decision and what is a scientific proof. In the case of a political decision, the unanimity and the number of voters are essential ingredients in weighting the validity of the decision taken. At contrary science has nothing to do with neither unanimity nor number of voters. Science policy does as well choices for funding but not science itself. One might recall that consensus of scientists regarding erroneous "truths" has often been used to oppose the acceptance of genuine new discoveries. A scientific proof can be discard by the scientific community for some times as with the famous examples of Galileo and Einstein.

Hence if one insists so much on the very broad consensus backing the "scientific proof" of human guilt for global warming, that in itself proves that the asserted "proof" is absent.  One must be very clear about this matter. At present, contrary to what has taken place during recent years, there exists no scientific certainty about human guilt concerning the global warming that.  There is only the strong conviction of thousands of scientists that it is so.  

This is not a negligible matter in putting priorities in the research objectives but it should not in any case be an argument to forbid parallel research in other directions. The debate must stay wide open within the community of climatologists.  The matter is simply not yet resolved scientifically, even if politically it appears to be. 

To make the issue at stake more precise, imagine that, for some incredible reason, tomorrow 10,000 physicists from all over the world unanimously voted that gravity does not existÑevidently such a vote would not alter the reality of gravity's existence by one iota.  On the other hand, such a vote could convince millions of people that they could safely jump off a high place, and consequently they would die, dismembered, on the ground.  That is the root of the present danger rooted in the positioning of a truth pretended scientific. 

It may be necessary to reassess how a scientific is presented. If one has some novel proof of a phenomenon (which, unlike mathematical proof, reposes essentially on an overwhelming accumulation of evidence and repeated experiments), one simply says or writes, "X et al. have demonstrated (or proved) thatÉ".  One never says, "the whole world scientific community, united in conclave, have unanimously decided thatÉ". It is in the domains of politics and sociology that consensus is accepted in order to justify a choice, precisely because there does not exist the possibility of proof (or incontrovertible accumulation of evidence) in favour of that choice.

Thus, the position of the scientistsÑto be precise those belonging to the community of climatologists, which has created unitary organs speaking with a single voice and engaging in political lobbying, and with the mediaÑthat are securing for themselves larger and larger financial subsidies, is particularly disturbing from the viewpoint of the free development of research.  The effective elimination from this community of "dissidents", indeed there are some, is the first cause for alarm.

\section{When the charge of the proof is up to the defendants}

Serious climatologists will recognize that consensus does not as such establish a proof and that doubt is always possible. But in return for this scientific based opening and in order to discard non sense claims, they will state that to get a credible doubt, it must be sustained by a proof or overwhelming solid evidence in favour of the non-guilt of mankind.  

And it must be clearly recognized that up until now, such proof cannot be given.  There exists no proof to innocent mankind. But here stands a fallacious reversal of what should be proved indeed. It is not the duty of the sceptics to have to bring a proof of whatever it is about which they are sceptical as long as they are not stating anything but their doubt about some claimed truth.  Rather, it is up to the scientists making the new assertion who must bring the corresponding proof, in this case of human guilt.

The terms of the debate have been inverted.  Guilt has been erected as the truth, and it is up to the defendants of the opposite view to bring proof of the absence of guilt.  This is an absurd trap in which to fall, and which distorts the entire debate.  This adroit deception has a pernicious effect. The respective roles of the opponents have been surreptitiously inverted, and all further real inquiry into the matter is now subject to a barrier in the shape of an automatic accusation of superfluity. Man has been declared guilty simply because, at the present time, no other bearer of guilt has been found, and as mentioned above there are moreover some superficially attractive reasons for ascribing guilt to him.

\section{The misleading use of probability}

Those climatologists convinced about global warning, in order to remain rigorous, stipulate that they are certain to a degree of $90\%$ concerning human guiltÑbut quickly add that $90\%$ is essentially the same as $100\%$, and emphasize that it would be almost criminal to wait for having actually reached a certainty of $100\%$ before acting.  This reasoning is offered as an elementary application of the so-called precautionary principle, and vaccination against infectious disease is frequently given as an example of the application of this principle.  

Unfortunately, there is no link between respective problem of vaccination and global warming. The use of a probabilistic concept in this later context leads to a serious and dangerous confusion.  The use of the notion of probability in order to evaluate a risk is based on the existence of a collection of identical alternative events, the realization of any one of which is largely random.  The probability of meeting an infected person, and of being infected at the meeting, is an example of a situation to which probabilistic concepts may be legitimately applied.  

The evaluation of the risk is only reliable when the statistics describing the event (in this case, infection) are sufficiently large.  Probability theory then allows to calculate what could be the result of the event of meeting somebody by chance.  Thus one can legitimately talk about the probability of being infected in a certain region, or indeed of the probability of winning the lottery.  In the case of the risk of infection, if the vaccine exists and has no undesirable side-effects, there is likely to be a good case for vaccinating oneself, even if the probability of being infected is much less than 90%.

On the contrary, to use the notion of probability in order to define the degree of confidence in the diagnosis of a unique problem may lead to dramatic errors.  In order to discover the truth about a specific unique problem, one has to somehow aggregate a large number of indications, many of which are very different from each other, each one revealing only one part of the overall truth.  Unlike the repetition of the same event, these different indications have very different statistical weights.  Some seem major, other minor.  One can gather a very large number of them, all pointing in the same direction (or perhaps not).  Progressively, a truth is apprehended in accord with all the available indications, but without necessarily being the truth.  There is no question here of a mathematical proof, nor of a unique and incontrovertible relation of cause and effect.

Until such proofÑor incontrovertible demonstrationÑhas been accomplished, some new indication found from some previously unsuspected or not investigated source has the potential to annihilate the entire conviction constructed up to that point, and to itself form the basis of the definitive establishment of the real truth.  The example of a person accused of a crime well illustrates the subtlety of the process of proof (in the non-mathematical sense) of guilt.  One may possess $99\%$ of the indications, yet a single additional fact whose veracity is not in doubt can, at the last minute, exonerate the accused person.  Each case is unique.  It is meaningless to apply statistics in such cases, and to attempt to do so leads to dangerous arbitrariness. Numerous judicial errors have resulted from this fallacy.

In the case of a political diagnosis of a unique situation, choice is made according to a conviction established on the basis of a certain number of indications, and not by a 'proof'.  The Bush administration was persuaded, to very high degree, of the presence of weapons of mass destruction in IraqÑyet the conviction was wrong.  What happened thereafter is well known.  This does not of course imply that every strong conviction is necessarily wrong.

In the case of the climate, we are facing a unique situationÑthat of the Earth. No statistics are possible, and hence to get as close as we can to incontrovertible demonstration is indispensable in order to avoid committing an error with irreparable consequences. If decisions are to be taken, at the present status of the scientific knowledge, they must be taken as political choices, and not as imposed by science.

\section{What is the scientific ground for human guilt?}

Three empirical facts, the increase of the global temperature of the planet, the increase in the carbon dioxide concentration in the atmosphere, and an increase in the production of carbon dioxide by humanity form the main ground on which thousands of climatologists assert the pretended undeniable conclusion of human guilt.

By comparing the graphs of global temperature with time, and of the increase of the quantity of carbon dioxide in the atmosphere with time, they postulate a cause-effect correlation between the two phenomena, which occur simultaneously.  But to correlate them in a unique relation of cause and effect is an erroneous simplification that leads to premature judgments.  The two effects may influence each other reciprocally, and they could also be produced separately by other independent factors that cannot be individually identified in the extremely complex global context of climate, which is still far from being understood.

On the basis of this postulate, climatologists have constructed models capable of reproducing the climate in the past, and then, on the basis of numerical simulations carried out on the computer, they can run the models into the future and make predictions.  Now, these models are intrinsically mere approximations to reality, but they are not themselves the reality.  How is it possible to be certain not to have neglected some factor, considered to be insignificant today, but which tomorrow may turn out to be essential in the evolution of the climate?  Models promoted 15 years ago are not the same as the models in use today and the current models will in turn also become obsolete.

While the use of model is a fantastic tool for scientific investigation, it should be always emphasized that any model contains an enormous dose of uncertainty that depends on the knowledge available at the moment of its construction. Hence, inevitably, a new discovery can at any instant invalidate the model.  A modelÑI construct myself models in my workÑshould serve to orient research, but not become a substitute for the reality that it attempts to describe.

In addition, one should distinguish between precise results established reasonably reliably for a local set of circumstances, and their generalization to a global context, especially when this context is constructed on the basis of models that by their very nature embody an a priori vision of the phenomenon and whose only justification is their ability to reproduce when simulated on a computer a certain number of empirically established results. The current models used in climatology may turn not to be false, they even may be valid, but they are not the reality. Therefore their predictions should be considered always with caution and a part of doubt. In particular before indicating the way what political and economical decisions should be implemented at the world scale.

\section{Why not reducing carbon dioxide emissions anyhow?}

People could argue that there is nothing bad, on the contrary, to reduce drastically the anthropogenic carbon dioxide emissions in the framework of combating waste and pollution.  It is true but the strategic question is how such a reduction will be implemented. 

It would be an enormous error to accede to a unique system of thought such as that currently emerging from the combination of political demands for the reduction of carbon dioxide emissions with antiscientific ideology calling for a 'return to nature', advocating a halt to development, and a moratorium on investment in technology and expenditure on scientific research.

In case the current climate changes have natural causes, focusing our entire efforts on a drastic reduction of anthropogenic carbon dioxide emissions, implying a suppression of our advanced technologies, could leave as defenceless in the face of a newly hostile nature, and could simply accelerate the disappearance of the human raceÑquite the opposite of the intended outcome.  On the contrary, it would appear to be indispensable to intensify researchÑscientific research aimed at understanding our universe betterÑin all directions in order to assist mankind to adapt himself to and protect himself from global warming.  If the rate of global warming turns out to be as rapid as is sometimes forecast, the development of new technology and new scientific understanding will be our only chance of survival. 

Therefore the reduction of pollution should be undertaken while optimizing human development and rationalizing our energy consumption without needlessly destroying our mode of life, could become a trigger to substantial technological innovations. To make a mistake by prematurely selecting human development as the cause of global warming and drastically breaking it could thereby be a fatal mistake for us as a species.  On the other hand, to make a mistake by prematurely asserting that global warming is natural will at least allow us to formulate appropriate remedial actions, regardless of the root cause, i.e. even if it is anthropogenic.  

Major climate changes have taken place on Earth in the past and will doubtless happen again in the future, accompanied a t each time by the disappearance of tens of thousands of species, and without any human intervention at all. That will happen with certainty in the future, so better to start to be ready for it.

\section{The social warming may turn worse than the global warming}

To render effective the study of the climate and the associated possibility of actions, one must also understand the dynamics of human nature in the face of collective fears. Otherwise a collective fear may produce a social instability, whose immediate consequences would be more dramatic than the ones eventually caused by global warming. Several advance warnings of some "social heating" can already be identified. Although they are minor, they might well be the first indices of a more significant collective crystallisation to come. The past history of human societies in crisis should be revisited in this context.

One first step could be the creation of a "world observatory" that would dispassionately list and study the collective fears appearing all over the planet, whether they are baseless or not. It includes the loss of climatic regularity, the technical development in general, terrorism and globalization.

Through the work of this observatory one could perhaps avoid needless social catastrophe, driven by fear, at the highest levels of political institutions. Some kind of parapets would have to be invented to prevent the coalescence of archaic, yet legitimate, fears of the majority of the population with the social intentions of an active minority promulgating for instance sacrificial expiation.  Available data on the history of human fears and their consequences could at least prevent a re-enactment of some appalling scenarios of the past.

\section{The priority against global warming}

The threats and current disturbances created by modifications of terrestrial climate make urgent the understanding of what is really going on, to allow eventual actions to curb them. And the only chance to achieve such a goal is to cool down the debate global warming by re-placing it in its natural setting, that of scientific debate.  

It is at this stage more than legitimate to raise doubt among the public, politicians and scientists, concerning a matter that should properly be the subject of dispassionate scientific research. To cast such a doubt should not in any way be taken as a tentative to thwart the reduction of pollution and the waste of natural resources. Pollution, global warming and globalization are different problems, though certainly connected, and one should take action on each front separately.  Our survival will ultimately depend on it.

In particular, if global warming is natural the situation will become much more painful than expected.  In that case there exists no guarantee that we can ultimately do anything about it. Moreover, even the elementary steps to be taken to protect ourselves, are not clearly definable, which in itself engenders an intolerable existential anguish. But to use a scapegoat to calm down such an unbearable stress would be not only dangerous for the near future, but will jeopardize the future.

\section{Conclusion}

To sum up above analysis of the social and human  aspects of global warming, most caution should be taken to prevent opportunistic politicians, more and more numerous, to subscribe to the proposed temptation of a sacrifice frame in order to reinforce their power by canalizing these archaic fears that are re-emerging.  Let us keep in mind that in a paroxysm crisis of fear, opinions can be activated very quickly among millions of mobilized citizens, ready to act in the same direction, against the same enemy: it then enough to designate it.  

Such kind of phenomena should be studied within the new emerging field of sociophysics, in particular the dynamics of minority opinion spreading and the rumor propagation [6, 7, 8].

%%%%%%%%%%%%%

%%%%%%%%%%%%%%%%
 \end{document}